\documentclass{revtex4}
\usepackage{graphicx}

\begin{document}

\title{Masses and radii of neutron and quark stars}

\author{Alessandro Drago}
\affiliation{
Dip. Fisica - Universit{\`a} di Ferrara and INFN - Sez. Ferrara, 44100 Ferrara, Italy}

\author{Andrea Lavagno}
\affiliation{Dip. Fisica - Politecnico di Torino and INFN - Sez. Torino, 10129 Torino, Italy}

\begin{abstract}
We discuss new limits on masses and radii of compact stars and we conclude that 
they can be interpreted as an indication of the existence of two classes of
stars: ``normal'' compact stars and ``ultra-compact'' stars. We estimate the 
critical mass at which the first configuration collapses into the second.
\end{abstract}

\maketitle

\"{O}zel~\cite{oz1} in her re-analysis of EXO 0748-676 observational data came to the conclusion that 
all soft equations of state of neutron star matter are ruled out. On the other hand, \"{O}ezel et 
al. \cite{oz2} re-analysing the data of 4U 1608-52, 4U 1820-30 and EXO 1745-248 concludes that 
only a very soft equation of state is consistent with the extracted mass-radius relations. 
Since the same technique is adopted in both papers, there is an apparent contradiction. 
Here we claim that the contrasting conclusions can be reconciled if  
``normal'' compact stars, made entirely or partially of 
nucleons, can exist at the same time as more tightly bounded configurations made of 
``exotic'' matter. The ultra-compact configuration can be a star composed partially or totally of 
quarks~\cite{scher,berez,blaschke}. 

The limits derived in Ref.~\cite{oz1}, 
taken by themselves, can be satisfied by a purely nucleonic star, but they can also 
be fulfilled by a hybrid or a quark star~\cite{nat}.  Instead, when the limits derived in 
Refs.~\cite{oz1,oz2} are 
discussed together, the number of possible choices drastically reduces, since a common scenario 
must explain all data. An example is provided in the Figure: using the analysis done in Ref.~\cite{dra} we 
interpret EXO 0748-676 as a hadronic star and the three objects discussed in \cite{oz2} as quark or 
hybrid stars. The purely hadronic configuration is built using a relativistic hadronic model 
incorporating hyperons~\cite{gm}, while the hybrid star configuration is obtained interpolating (via Gibbs 
equations) between the same hadronic model and a MIT-bag model in which quarks form a 
Colour-Flavour-Locked (CFL) phase~\cite{alf}. The quark star configuration is again based on a CFL
phase, but using a larger value for the superconducting gap \cite{dra}.
It is interesting to remark that the above configurations were built to 
satisfy other and independent astrophysical constraints than those discussed in Refs.~\cite{oz1,oz2}. 

A crucial point in this scenario is that there is a critical baryonic mass for the 
``normal'' configuration above which the star becomes so meta-stable that it collapses into the more 
compact configuration~\cite{berez,bom}. By considering both the results of Ref.~\cite{oz1} and of Ref.~\cite{oz2} 
it is possible to put 
tight limits to the critical baryonic mass: it must be larger than the baryonic mass of 
the ``normal'' configuration (the 
hadronic one in our case) and lower than the baryonic mass of the compact 
configuration. Using the 
limits of Refs.~\cite{oz1,oz2} at 1-$\sigma$ we conclude that 
the critical baryonic mass is about $(1.9 \pm 0.1) M_\odot$, corresponding to 
$(1.8 \pm 0.1) M_\odot$ for the gravitational mass of the ``normal'' configuration and to 
$(1.6 \pm 0.1) M_\odot$ for the ultra-compact configurations. 

The interpretation we are suggesting poses questions and constraints to the structure of the 
ultra-compact configuration: 1) the critical baryonic mass is large. It is not easy to satisfy this constraint
if the ultra-compact configuration is a hybrid star:
in models describing deconfinement as a first order transition the value of the surface tension
would have to be so large that a mixed-phase could not form~\cite{berez,bom}, while models based
on the ``third family'' scenario~\cite{scher,blaschke} have difficulties in reaching large
values for the mass of the ``normal'' configuration;
2) the possible co-existence of quark 
stars and of ``normal'' stars has been recently reconsidered in Ref.\cite{bau}, showing that it is not forbidden. 
On the other hand, a quark star entirely made of CFL matter is unlikely if it is 
rapidly rotating (as for 4U 1608-52), since its viscosity is negligible. A quark star made at least in 
part of normal or of 2SC quarks could solve this problem.

Finally, the existence of both a ``normal'' and of a ultra-compact 
configuration opens the possibility of a huge energy release, of the order 
of a few $10^{53}$ erg, associated with the transition 
from one configuration to the other. That energy can help en-powering explosive astrophysical 
phenomena~\cite{cheng,berez}.

\begin{figure}
\begin{center}

\includegraphics[height=10cm,width=15cm]{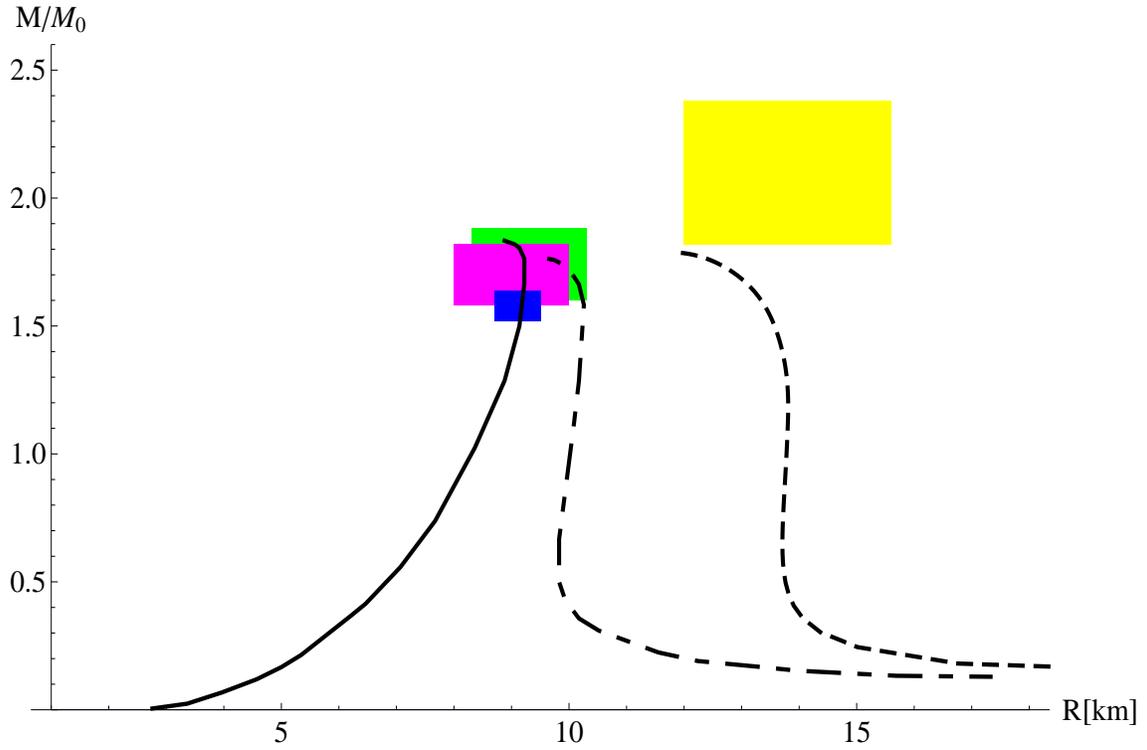}
\end{center}
\caption{Mass-radius plane with observational limits and representative
theoretical curves. Solid line indicates CFL quark stars; 
dot-dashed line, CFL hybrid stars; dashed line, hadronic stars
(same lines as in Ref.~\cite{dra}).
The yellow box corresponds to 1-$\sigma$ confidence contours of EXO 0748-676 \cite{oz1},
the green, magenta and blue boxes correspond to the confidence contours of 4U 1608-248,
EXO 1745-248 and 4U 1820-30, respectively (see Ref.~\cite{oz2} and references therein).}
\end{figure}

\end{document}